%% file: 00-main.tex
\documentclass[sigconf,screen]{acmart}
\pdfoutput=1
\usepackage[utf8]{inputenc}
\usepackage[T1]{fontenc}
\usepackage[normalem]{ulem}
\usepackage{amsmath}
\usepackage{bbm}
\usepackage{xcolor}
\usepackage{natbib}
\usepackage{graphicx}
\usepackage[skip=-1pt]{caption}
\usepackage{subcaption}
\usepackage{amsmath}
\usepackage{makecell}
\usepackage{amsfonts}
\usepackage{hyperref}
\usepackage[inline]{enumitem}
\usepackage{booktabs}
\usepackage{multirow}
\usepackage{mathtools}
\usepackage{multicol}
\usepackage{amsmath, bm}
\usepackage{stmaryrd}
\usepackage{longtable}
\usepackage{autobreak}
\usepackage{breqn}
\usepackage{titlesec}
\usepackage{scalerel}
\usepackage{tabu}
\usepackage[ruled,vlined,english,onelanguage]{algorithm2e}
\usepackage{cleveref}

\newcommand{\eg}{\emph{e.g.}}

\newcommand{\vs}{\emph{vs. }}

\AtBeginDocument{%
  \providecommand\BibTeX{{%
    \normalfont B\kern-0.5em{\scshape i\kern-0.25em b}\kern-0.8em\TeX}}}

\setcopyright{none}
\copyrightyear{2022}
\acmYear{2022}

\acmConference[FAccTRec '22]{The 5th FAccTRec Workshop on Responsible Recommendation at RecSys 2022}{September 18--23, 2022}{Seattle, WA \& Online}
\acmBooktitle{FAccTRec '22: The 5th FAccTRec Workshop on Responsible Recommendation at RecSys 2022, September 18--23, 2022, Seattle, WA \& Online} 

\begin{document}
\fancyhead{}

\title{Ethical and Social Considerations in Automatic Expert Identification and People Recommendation in Organizational Knowledge Management Systems}

\author{Ida Larsen-Ledet}
\email{t-idal@microsoft.com}
\affiliation{%
  \institution{Microsoft Research}
  \streetaddress{21 Station Road}
  \city{Cambridge}
  \country{United Kingdom}
}

\author{Bhaskar Mitra}
\email{bmitra@microsoft.com}
\affiliation{%
  \institution{Microsoft Research}
  \streetaddress{6795 Rue Marconi}
  \city{Montreal}
  \country{Canada}
}

\author{Si\^{a}n Lindley}
\email{sianl@microsoft.com}
\affiliation{%
  \institution{Microsoft Research}
  \streetaddress{21 Station Road}
  \city{Cambridge}
  \country{United Kingdom}
}

\input{01-abstract}


\maketitle

\input{02-intro}
\input{03-harms}
\input{04-conclusion}
\balance
\bibliographystyle{ACM-Reference-Format}
\bibliography{references}
\end{document}

%% file: 01-abstract.tex
\begin{abstract}
Organizational knowledge bases are moving from passive archives to active entities in the flow of people's work. We are seeing machine learning used to enable systems that both collect and surface information as people are working, making it possible to bring out connections between people and content that were previously much less visible in order to automatically identify and highlight experts on a given topic. When these knowledge bases begin to actively bring attention to people and the content they work on, especially as that work is still ongoing, we run into important challenges at the intersection of work and the social. While such systems have the potential to make certain parts of people's work more productive or enjoyable, they may also introduce new workloads, for instance by putting people in the role of experts for others to reach out to. And these knowledge bases can also have profound social consequences by changing what parts of work are visible and, therefore, acknowledged. We pose a number of open questions that warrant attention and engagement across industry and academia. Addressing these questions is an essential step in ensuring that the future of work becomes a good future for those doing the work. With this position paper, we wish to enter into the cross-disciplinary discussion we believe is required to tackle the challenge of developing recommender systems that respect social values.
\end{abstract}

%% file: 02-intro.tex
\section{Introduction}
\label{sec:intro}

Work in large organizations generates vast amounts of information that bring challenges to knowledge management and dissemination. Recent developments have seen machine learning (ML) successfully applied in the automatic construction of knowledge bases that aim to tackle these problems~\cite{winn2021enterprise}, and products like Viva Topics~\cite{VivaTopics} are bringing these capabilities to workplaces. Viva Topics identifies organizational ``topics'', associates them with organization members and resources, and surfaces these details via office software. As an example, the topic ``Accessibility'' might be associated with a company’s accessibility guidelines and the contact details of an in-house expert on accessible design, and then highlighted on relevant SharePoint pages.
ML-powered systems like these enable automatic creation, sharing, and maintenance of organizational knowledge and have the potential to (re-)shape and surface that knowledge in the flow of people’s work~\cite{Wilkins2020}. Looking ahead, knowledge bases may become an increasingly (pro-)active part of work, by recommending documents or experts that are relevant to the task a person is currently working on. In doing so, knowledge recommendations will represent people and their work, thereby influencing the visibility~\cite{Bowker2000, larsen-ledet2022} of their efforts and, potentially, their opportunities for impact~\cite{Lindley2021}. This means that, in addition to known fairness challenges of recommender systems~\cite[\eg,][]{ekstrand2022fairness, olteanu2021facts, burke2021preface, ekstrand2018all, Neophytou2021}, we will need to address issues particular to organizational knowledge management systems and the ways knowledge bases mediate and represent work and, hence, people.

In this position paper, we outline a set of concerns regarding issues of representation in the deployment of recommender systems that surface people and their work within their own organizations. These concerns hint at sensitivities particular to the domain of organizational knowledge bases; these are sensitivities that we and others will need to develop for these kinds of systems. We wish to start a conversation with the community, in which these social and ethical concerns can be addressed in a cross-disciplinary setting.

%% file: 03-harms.tex
\section{Consequences of recommending people}
\label{sec:harms}
The design of online recommender systems raises several concerns related to fairness and transparency~\citep{ekstrand2022fairness, olteanu2021facts, burke2021preface, ekstrand2018all, Neophytou2021}.
The stakes are high when the subject of recommendations are people, especially when these systems fail to be meaningfully transparent about the context in which these people are recommended~\citep{li2022exposing} and to provide effective recourse to people in case of problematic recommendations.
A person may be recommended based on their self-declared attributes, preferences, and expertise---\eg, candidate recommendations for jobs~\citep{geyik2018talent} and online dating applications~\citep{zheng2018fairness}---or their attributes and expertise may be automatically inferred from contents they have authored and other traces of their relevant activities---\eg, in an organizational setting~\citep{VivaTopics}.
The latter approach deserves additional scrutiny, not just with respect to the correctness of inferred expertise but also the implications of potential epistemic injustices~\cite{fricker2007epistemic} and how the system may influence work practices.
Extensive critical analysis is necessary to identify these biases and harms in the context of organizational knowledge and expertise modeling that uses machine learning.
In particular, we should ask ourselves the question: \emph{What are relevant forms of ``social and human impact'' in the context of organizational recommender systems?}
Here, we enumerate some of the risks that we already anticipate along with open questions to set off a discussion to help the community approach those risks.

\paragraph{Erasure.}
Erasure refers to practices and collective actions that render\,certain\,individuals\,and\,groups\,invisible~\citep{sehgal2016fighting},\,resulting\,in\,their marginalization and dehumanization (representational harm~\cite{crawford2017trouble}).
An organizational knowledge management system that fails to adequately recognize different types of expertise or 
the expertise of certain individuals or groups contributes to the erasure of their labor.
Such erasure could limit a person's opportunities for recognition and, in the long run, career advancement and work satisfaction (allocative harm~\cite{crawford2017trouble}).
There are several mechanisms by which such erasure may actualize in ML-enabled organizational knowledge management systems.
For instance, if expertise is inferred from authored content, the system may over-emphasize areas and types of work that produce more textual output, and may misattribute or misrepresent expertise when the person doing the work is not the one who documents it, or when the outputs of a person's work are not as easily mined or classified~\cite{Bowker2000}.
Erasure can also be caused by abstractions~\cite{o2016weapons, jacobs2021measurement} and demographic data gaps~\cite{perez2019invisible} that skew the system towards expertise indicators historically associated with the demographics dominating the training data (typically white able-bodied cis-gendered men~\cite{perez2019invisible}).
This prompts us to suggest that, in addition to asking ``what content and data is useful," we should be asking: \emph{``What content and data is needed to adequately represent expertise?''}

\paragraph{Bias amplification.}
Biases and harms are not just consequences of system design but are further compounded by the organizational practices in which these systems are situated and how users interact with them.
For example, if the system presents its recommendation as a static list of people ranked by their expertise on a topic, small differences in estimated expertise may translate into disparately different levels of exposure, as presentation bias causes people to over-prioritize those at the top of the list~\citep{singh2018fairness, diaz2020evaluating, wu2022joint}.
This could have a self-reinforcing effect, in which ``top experts'' have more interactions and encounter more opportunities that help them further develop their expertise, leading to a rich-gets-richer effect and ``economics of superstars''~\cite{rosen1981economics, baranchuk2011economics}.
Conversely, being recognized as an expert on a specific topic may put additional labor onto a person, reducing the time they have to develop or diversify their expertise.
To create systems that are resilient to distortions of visibility and work practices, we need to ask: \emph{What distortions may occur in organizational knowledge bases and what will it mean to be resilient to them?} Beyond that, we ought to explore if ML-powered knowledge bases give us the potential to strive for more than resilience, such as for knowledge bases to actively combat distortion. For instance, \emph{what is a good balance between providing tailored recommendations and surfacing content that diverges from historical patterns and expands the individual's outlook on organizational resources?}


\paragraph{Creating perverse incentives.}
If being recognized as an expert leads to increased opportunities and rewards, people might respond by adapting their practices to ensure visibility and acknowledgment of their work.
This risks producing a work culture focused on performativity rather than enduring contributions and long-term value~\cite{Linehan2011}, and may eventually make the applied measures of expertise less meaningful~\cite{goodhart1984problems}.
For example, people may feel incentivized to produce outputs of a form that the system picks up on even though that may not be what makes the most sense in practical terms---one might imagine designers producing high-fidelity digital prototypes in situations where a rough paper mock-up would be more suitable.
These dynamics may also encourage and normalize uncompensated labor extraction from employees, for them to ensure that their work is seen by the system and, by extension, their organization.
For knowledge bases to foster healthy dynamics, we need to challenge existing designs by asking: \emph{What gaps does the current design introduce or reproduce in terms of expertise representation?} In order to examine this question in a given organizational context, we may begin by asking: \emph{What kinds of expertise exist in the organization? What objects are these different kinds of expertise tied to?}
And, by extension: \emph{What forms of knowledge production are, indirectly, encouraged or discouraged by existing systems and practices?} Finally, with a more generative outlook~\cite{Beaudouin-Lafon2021}: \emph{How can knowledge bases encourage the development and nurturing of diverse expertise?}

\paragraph{Knowledge production \vs knowledge extraction.}
As we have hinted at above, there is more at stake than fairness:
Active knowledge bases not only \emph{represent} extracted knowledge but also \emph{shift} the way knowledge production happens, by mediating interactions between people and influencing work through suggestions and surfaced content.
This underscores the need to understand local practices of knowing~\cite{Orlikowski2002}, so as to be deliberate in how the system will slot into and alter those practices.
In most organizations, the identification of expertise is inherently a social and situated process~\cite{Oreskes2021}, whereas classification systems tend to codify expertise into rigid structures that do not always exhibit the same contextual flexibility~\cite{Bowker2000}.
This leads us to ask how these systems can leave room for the situated sense-making and consensus building that helps people trust knowledge and expertise~\cite{Oreskes2021}, through questions such as:
\emph{What would ``sociality'' mean in these kinds of systems? What will consensus building and trust look like if automated knowledge bases and recommender systems become key infrastructures and mediators (or what do we want them to look like)?}
Furthermore, to ensure that the \emph{extractive} process of constructing organizational knowledge bases does not adversely affect knowledge \emph{production} within the organization, we also need to address how to make systems that respect people's autonomy~\cite{Bowker2000}. This hints at questions about workplace monitoring and control of one's data (see below), but more directly related to knowledge production, we may ask: \emph{What additional work does increased or altered visibility impose on people}?~\cite{larsen-ledet2022} \emph{Who is work shifted \emph{from} and \emph{onto}, and do those who are doing the work also benefit (sufficiently) from it?}~\cite{Lampinen2022}


\paragraph{Organizational knowledge \vs control and organizational power.}
Our final critique on the application of machine learning for extracting and surfacing organizational knowledge is that, unless deliberately countered, these systems will reflect existing power structures, with the risk that existing power imbalances manifest in
abuse such as workplace surveillance~\citep{teachout2022boss}. This prompts questions like the following: \emph{What kinds of accountability does the visibility brought about by these systems impose on the people being represented through the system? In what ways might this accountability result in abuse?} One topic to address here is the relationship between control and the temporality of visibility:
Organizational knowledge objects are not static. Documents and other content produced by employees have a life cycle, moving through stages from work in progress, to finished, to approved, to outdated, etc. The content picked up by an automated knowledge base will often be work in progress and, as such, knowledge objects may be made available to others, or even actively distributed, before they are ``ready,'' thereby taking away the authors' control over how they appear to colleagues through the objects of their work~\cite{Larsen-Ledet:2019}. From a design perspective, it is worth asking: \emph{How can automated knowledge bases be attentive to context such as timing?}

%% file: 04-conclusion.tex
\section{Beginning the conversation}
\label{sec:conclusion}
These kinds of systems are still in their early stages, but as researchers and practitioners we have a responsibility to anticipate and mitigate risks like the ones outlined here, as early as possible~\cite{Parvin2020}.
Any meaningful mitigation strategy necessitates causal attribution of the sources of said risks---\eg,
\begin{enumerate*}[label=(\roman*)]
    \item the framing of the problem that the machine learning system should solve may itself encode assumptions, such as that expertise can be inferred from authored content, raising questions of construct validity~\cite{jacobs2021measurement},
    \item datasets collected under unjust social biases and power inequities may contribute to erasure and unfair outcomes,
    \item the choice of optimization objectives, such as mean performance, may overlook poor recommendation quality for minority groups of users,
    \item the models themselves, for example when they do not consider causality, may result in stereotyping, and
    \item presentation bias in how the system displays the recommendation results, such as a static ranked list, may further amplify inequities in exposure.
\end{enumerate*}
Here, we have only begun to speculate about potential issues, but even this demonstrates the urgency of more thoroughly identifying pitfalls. In other words, while predictive machine learning may, indeed, have much to offer organizational knowledge management, we must remain cognizant that no vision for the ``future of work'' (a seemingly pervasive phrase across industry, business, and academia,\footnote{See, e.g., Microsoft (\url{www.microsoft.com/research/theme/future-of-work}), Forbes (\url{www.forbes.com/feature/future-of-work}), OECD (\url{www.oecd.org/future-of-work}), and Harvard University (\url{www.harvard.edu/the-future-of-work}).} to name a few areas) is complete without adequate emphasis on \emph{the future of workers}, which in turn may emphasize the need for participatory design~\citep{Bodker2018} and protection for worker rights and the labor movement.
Some of the questions we have posed in the section above pertain primarily to one of the sources mentioned here, while others can be asked from multiple angles. For instance, being attentive to context (the final question posed) may be a question of what signals are included in the data or how they are used in the model; but it may also be a question of what we assume different contextual signals to mean or how we choose to design the user interface based on them.
At a more political level we may pose the questions: \emph{Who should influence and control the construction of organizational recommender systems at various stages, from design to adoption and use? Who owns the knowledge extracted from activities at work? What does it do to a person's ``ownership'' of a process if that process is modelled and made replicable? What would be pros and cons, from an end-user perspective, of being given control over how one's work data is interpreted and when it is used?} The questions are many, and we are only just beginning the conversation.